\documentclass[a4paper]{jpconf}
\usepackage{graphicx,feynmf}

\newcommand{\abb}{Fig.~\ref}
\newcommand{\fig}{\abb}

\newcommand{\TeV}{\,\mbox{Te\kern-0.2exV}}
\newcommand{\GeV}{\,\mbox{Ge\kern-0.2exV}}
\newcommand{\mGeV}{\,\mathrm{Ge\kern-0.2exV}}
\newcommand{\MeV}{\,\mbox{Me\kern-0.2exV}}
\newcommand{\keV}{\,\mbox{ke\kern-0.2exV}}
\newcommand{\eV}{\,\mbox{e\kern-0.2exV}}
\newcommand{\km}{\,\mbox{km}}

\newcommand{\ipb}{\,\mbox{pb}^{-1}}

\newcommand{\pb}{\,\mbox{pb}}

\newcommand{\met}{\slash\!\!\!\!{E_T}}
\begin{document}
\title{D\O\  Top Quark Results and their Dependence on Successful Grid Computing%
}

\author{Daniel Wicke}

\address{Fermilab MS357, P.O. Box 500, Batavia, IL 60510\footnote{on  leave
    from Bergische Universit\"at Wuppertal}}

\ead{wicke@fnal.gov}

\begin{abstract}
The heaviest known Fermion particle --- the top quark --- was discovered at
Fermilab in the first run of the Tevatron in 1995. However, besides its mere
existence one needs to study its properties precisely in order to verify or
falsify the predictions of the Standard Model.
With the top quark's extremely high mass and short lifetime such measurements
probe yet unexplored regions of the theory and bring us closer to solving the
open fundamental questions of our universe of elementary
particles such as why three families of quarks and leptons exist and why
their masses differ so dramatically.

To perform these measurements hundreds of millions of recorded
proton-antiproton collisions must be reconstructed and filtered to extract the
few top quarks produced. Simulated background and signal events with full
detector response need to be generated and reconstructed to validate and
understand the results.
Since the start of the second run of the Tevatron the D\O\ collaboration
has brought Grid computing to its aid for the production of simulated
events. Data processing on the Grid has recently been added and thereby
enabled us to effectively triple the amount of data available with the highest
quality reconstruction methods.

We will present recent top quark results D\O\ obtained from these improved data
and explain how they benefited from the availability of computing resources on
the Grid.
\end{abstract}

\section{Introduction}

Elementary particle physics aims to find and describe the fundamental
building blocks of matter and the forces by which they interact.

'Normal' matter can be built out of only 3 elementary particles:
electrons and two types of quarks labeled $u$-quark and $d$-quark.
Electrons form the atomic shell. Triplets of quarks build the protons
and neutrons that form the atomic nucleus. Protons consist of two $u$-quarks 
and one $d$-quark; neutron consist of two $d$-quarks 
and one $u$-quark. A 4th particle, the neutrino, is needed to explain
radioactive $\beta$-decays.

The fundamental matter particles, the Fermions,
interact via four forces: Electromagnetism which
binds the electrons to the atomic core,
the strong nuclear force which binds the protons and neutrons within the
atomic nucleus, the weak nuclear force which mediates radioactive $\beta$-decays,
and gravitation (which is neglected in elementary particle physics).

A feature of the weak interaction (CP-violation) requires that the 
above four particle, which are called the first generation of fermions, are accompanied 
by two more generations of 4 fermions each. The fermions of the second and
third generation have quantum numbers identical to those of the first
generation, but higher masses. 

The final piece in the Standard Model is the mechanism of how elementary
particle can obtain mass. The symmetries of the theory prohibit explicit mass
terms in the fundamental equations. The so called Higgs mechanism
overcomes this problem by introducing spontaneous breaking of the electro-weak symmetry. 
Besides ``giving'' mass to  elementary particles
it also predicts the existence of a spin 0 particle, the as yet unobserved
Higgs boson.

  \begin{table}[tbp]
    \centering
\begin{tabular}{lllllll}
Leptons&$\nu_e$     &  electron neutrino & $\nu_\mu$     &  muon neutrino & $\nu_\tau$ & tau neutrino \\
&$e$         &  electron               & $\mu$       &muon              &$\tau$  &tauon\\
~\\
Quarks&$u$       &  up-quark & $c$       &  charm-quark&  $t$       &  top-quark\\
&$d$       &  down-quark&  $s$       &  strange-quark& $b$       &  bottom-quark
\end{tabular}
    \caption{Elementary particle building matter. Fermions of the Standard Model.}
    \label{tab:sm_fermions}
  \end{table}

  \begin{table}[tbp]
    \centering

\hfill Electromagnetism: Photon $\gamma$,\hfill Weak force: $Z,W^\pm$,\hfill
Strong force: Gluon $g$, \hfill Gravitation: Graviton\hfill
    
    \caption{Particles mediating fundamental forces. Vector bosons of the Standard Model.}
    \label{tab:sm_bosons}
  \end{table}

When the top quark was discovered in 1995 at Fermilab~\cite{Abe:1995hr,Abachi:1995iq} it completed the
set of quarks predicted by the Standard Model. Its mass was determined to be
30 times higher than that of the second heaviest fermion thereby being very
close to the electroweak symmetry breaking scale.

Within the Standard Model the properties of the top quark with the exception
of its mass are fully defined. The various possible extensions of the Standard
Model predict different variations of these properties.
The high mass of the top quark and its closeness to the electroweak symmetry
breaking scale has led to speculations that the top quark
plays a special role among the elementary particles which is not reflected in
the current Standard Model.
Measuring the properties of the top quark is therefore essential to
complete the verification of the Standard Model and to check the proposed
theories of new physics. 
%

In the following it will be described how such measurements are performed. 
Section \ref{sec:experiment} describes the Tevatron accelerator, the D\O\
experiment and how the obtained data are prepared for physics analysis.
In section \ref{sec:grid}  the tools needed to actually perform this
data preparation and to provide the resulting data to the physicists for
analysis are discussed.
Finally in sections \ref{sec:toppairxsec} to \ref{sec:topmass}
with the example of three important measurements it is explained how
recent D\O\ top quark results are obtained
emphasising in which way the analyses rely on  D\O's ability 
to utilise grid computing.

\section{Experiment}
\label{sec:experiment}
\subsection{The Fermilab Tevatron}
To produce top quarks (in pairs) the energy equivalent to (twice) the top
quarks mass needs to be concentrated into a volume which 
allows it to be to consumed in a fundamental reaction, ${\cal O}(1\,\mathrm{fm})$.

The only machine which is currently capable of achieving this is the Tevatron
proton-antiproton collider at Fermilab. The Tevatron is an accelerator
ring with a $7\km$ circumference. Protons and antiprotons are accelerated
around the ring in opposite
directions until they reach an energy of $980\GeV$. 
The beams of  protons and antiprotons are then brought to collision at
two interaction points both of which are equipped with detectors to record the
resulting events. The detectors are
CDF and D\O.

After an initial run in 1985--1995 the Tevatron has been upgraded
and is operating again  in  its so called Run~II since 2001.

\subsection{D\O\ Detector}
The detectors to record the collision events are built and operated by
international collaborations. The D\O-collaboration consists of around 670
physicists from 86 institutes in 19 countries on 4 continents.

Like all modern day  detectors in high energy particle physics D\O\ consists of
three main detection systems, which are placed in a cylindrical structure
around the beam pipe. 

The innermost system is the tracking system which is used to detect charged particles.
In D\O\ it consists of a high resolution silicon microstrip detector  
and a scintillating fiber tracker.
The complete tracking system is enclosed in a superconducting solenoidal
magnet creating a field of $2$T. This enables D\O\ to measure the momentum 
and the charge sign of the detected particles.
Efficient measurement of charged particles extends to pseudorapidities of
$\left|\eta\right|<3$, where pseudorapidity is a measure of azimuthal angle
with respect to the beam axis, $\eta=-\ln \tan \vartheta/2$.

The calorimeter surrounds the tracking system. It aims to measure
the energy of charged \em  and \em neutral particles by complete absorption.
The D\O\ calorimeter is built of liquid argon and uranium. It is separated
into a central part (CC) which covers $\left|\eta\right|<1$ and two end
calorimeters (EC) which cover to $\left|\eta\right|\simeq 4$.
By varying absorber thickness and materials the section closest to the
interaction point within each of the three calorimeter parts  
specialises on absorbing electrons and photons while only the
outer two section absorb the hadrons.
The distinction between the inner sections (the electromagnetic
calorimeter) and the outer two sections (the hadronic calorimeter) makes it
possible to distinguish
electrons and photons from hadronic particles.

Usually only muons and neutrinos escape these calorimeters.
To identify muons the calorimeters are surrounded by 3 layers of 
drift tubes, the so called muon chambers. A toroidal magnetic field between
the innermost two layers allows D\O\ to improve the measurement of muon momenta.
The presence of neutrinos has to be inferred from a transverse momentum imbalance.
%

In total D\O\ has around $1$ million readout channels. 
All signals produced in the various detector components are digitised 
and collected into an event record and stored to tape in files which typically
contain a few thousand events each.
At a data taking rate of $50\,\mathrm{Hz}$ and an  average event size of
$250\,\mathrm{kB}$ D\O\ writes 
$1.3\mathrm{MB}/\mathrm{s}$ to tape. The total amount of data recorded so far is
$\sim 400\mathrm{TB}$.

Before these data are used for physics analyses a set of common and compute
intensive reconstruction algorithms is applied to the raw data.
Signals from the tracker are passed through pattern recognition algorithms
which reconstruct the tracks of individual charged particles within the
detector and determine their charges and momenta;
Calorimeter cells are combined into jets of energy with various jet
reconstruction algorithms. From these then more global event properties like
e.g. the missing transverse energy are computed.

\section{Grid}
\label{sec:grid}
In order to handle its large amount of data and to serve them
to institutes around the world D\O\ uses grid technologies.
Since recently the grid is also used for the distribution of jobs required to perform the
central reconstruction and simulation tasks.

\subsection{SAM --- Sequential Access through Metadata}
SAM is D\O's data handling system~\cite{d0note3464}. It exploits the fact that events are
independent of each other and thus the order in which events are processed doesn't
matter.

Users request `datasets' instead of ordered lists of files.
SAM then optimises the order in which it presents the files to the
users to minimise the number of copy or tape operations.
In SAM each file has metadata describing its content.
These metadata are used to describe the datasets.

For derived files the metadata also contain information about the file's parents and
about the application and version it was produced with.
This information provides a complete book-keeping for any production operation.

Built on this data-handling system D\O\ has created a tiered infrastructure which
allows coherent data access from all over the globe.

\subsection{JIM --- Job Information Monitoring}
JIM aims to provide job submission to D\O's distributed resources integrated
with the SAM data-handling
system.
It is based on globus~\cite{globus} and
condor~\cite{condor-flock}. 
JIM also provides monitoring of remote (batch-)jobs. This monitoring
information is held in an XML database. A standard view to the information is
accessible via the Web.
The combination of SAM and JIM is called the
SamGrid~\cite{Baranovski:2003mr,Baranovski:2003wg,Baranovski:2003is}. 

\subsection{Application}
D\O\ is relying on the described capabilities in distributed computing to
perform all its generation of simulated events since the beginning of Run~II in
2001. Alone in the last year 80 million events with full detector response
simulation corresponding to $40$TB of data were generated and reconstructed remotely.
A stable and reliable data-handling system is required for this task.

In addition D\O\ is using distributed computing for re-reconstruction of data.
Re-reconstruction of data enables us to apply the most recent and most
advanced algorithm to data which has been reconstructed before with older software versions.
Improvements in the algorithms result from thorough investigation of the actual
detector performance.

In a first effort at the end of 2003 300 million data events were reprocessed from 
an intermediate data format. Significantly improved tracking algorithms and
improved tables of hot and dead calorimeter cells were applied.
$45$TB had to be read from tape and processed.
30\% of this effort was done at non-dedicated remote sites.
Following this effort the dataset available for this years publications using
the improved algorithms could be doubled.

\label{sect:d0repro}
Currently D\O\ is again reprocessing its full dataset. Improved 
calorimeter calibration is applied during this reconstruction.
As some of the information required for this computation is only available in
the original raw data this reprocessing is performed from raw data. The
reconstruction of $1$ billion events involves reading $250$TB from tape and
distributing them to the participating site.
All participating sites need the ability to access the central calibration
database either directly or through a local proxy server.
It was planned to perform the complete effort on remote sites as the D\O\
processing farm
at Fermilab, being busy with current data-taking, can only contribute to a small fraction.

In this effort the data-handling, the job distribution and the associated
book-keeping capabilities of SamGrid are used. It is expected that the current
reprocessing will double the dataset available for analysis
with upto date reconstruction algorithms during early 2006.

\section{Top pair production cross-section}
\label{sec:toppairxsec}
\begin{figure}[tbp]
\begin{minipage}[b]{0.48\textwidth}
  \null\hfill
  \begin{fmffile}{mf_s_chanel_ttbar}
    \begin{minipage}[c]{30mm} \footnotesize
    \setlength{\unitlength}{0.7mm}
    \begin{fmfgraph*}(35,30)
      \fmfstraight
      \fmfleft{em,ep}
      \fmfright{q,qb}
      \fmf{plain}{em,v,ep}
      \fmf{gluon,lab=\raisebox{-2mm}{$g$}}{v,vv}
      \fmf{fermion}{q,vv,qb}
      \fmflabel{$q$}{ep}
      \fmflabel{$\bar{q}$}{em}
      \fmflabel{$t$}{q}
      \fmflabel{$\bar{t}$}{qb}
    \end{fmfgraph*}
   \end{minipage}
  \end{fmffile}\hfill
  \begin{fmffile}{mf_t_chanel1_ttbar}
    \begin{minipage}[c]{30mm} \footnotesize
    \setlength{\unitlength}{0.7mm}
    \begin{fmfgraph*}(35,30)
      \fmfstraight
      \fmfleft{g1,g2}
      \fmfright{q,qb}
      \fmf{fermion}{q,v,vv,qb}
      \fmf{gluon}{g1,v}
      \fmf{gluon}{g2,vv}
      \fmflabel{$g$}{g1}
      \fmflabel{$g$}{g2}
      \fmflabel{$t$}{q}
      \fmflabel{$\bar{t}$}{qb}
    \end{fmfgraph*}
   \end{minipage}
  \end{fmffile}\hfill\null\\[4em]
  \null\hfill
  \begin{fmffile}{mf_t_chanel2_ttbar}
    \begin{minipage}[c]{30mm} \footnotesize
    \setlength{\unitlength}{0.7mm}
    \begin{fmfgraph*}(35,30)
      \fmfstraight
      \fmfleft{g1,g2}
      \fmfright{q,qb}
      \fmf{gluon}{g1,v}
      \fmf{gluon}{g2,vv}
      \fmf{phantom}{q,v,vv,qb}
      \fmffreeze
      \fmf{fermion}{q,vv,v,qb}
      \fmflabel{$g$}{g1}
      \fmflabel{$g$}{g2}
      \fmflabel{$t$}{q}
      \fmflabel{$\bar{t}$}{qb}
    \end{fmfgraph*}
   \end{minipage}
  \end{fmffile}\hfill
  \begin{fmffile}{mf_g_fusion_ttbar}
    \begin{minipage}[c]{30mm} \footnotesize
    \setlength{\unitlength}{0.7mm}
    \begin{fmfgraph*}(35,30)
      \fmfstraight
      \fmfleft{g1,g2}
      \fmfright{q,qb}
      \fmf{fermion}{q,vv,qb}
      \fmf{gluon}{g1,v,g2}
      \fmf{gluon}{v,vv}
      \fmflabel{$g$}{g1}
      \fmflabel{$g$}{g2}
      \fmflabel{$t$}{q}
      \fmflabel{$\bar{t}$}{qb}
    \end{fmfgraph*}
   \end{minipage}
  \end{fmffile}
\vspace*{2em}
\hfill\null
 \caption{Feynman diagrams contributing to top pair production. The $q\bar q$
   annihilation (upper left) dominates at Tevatron energies. The
    three diagrams with gluons in the initial state
    are expected to contribute  $15\%$.\label{fig:ttbar_feynman}}
\end{minipage}
\hfill
\begin{minipage}[b]{0.48\textwidth}
  \centering
  \includegraphics[width=\textwidth]{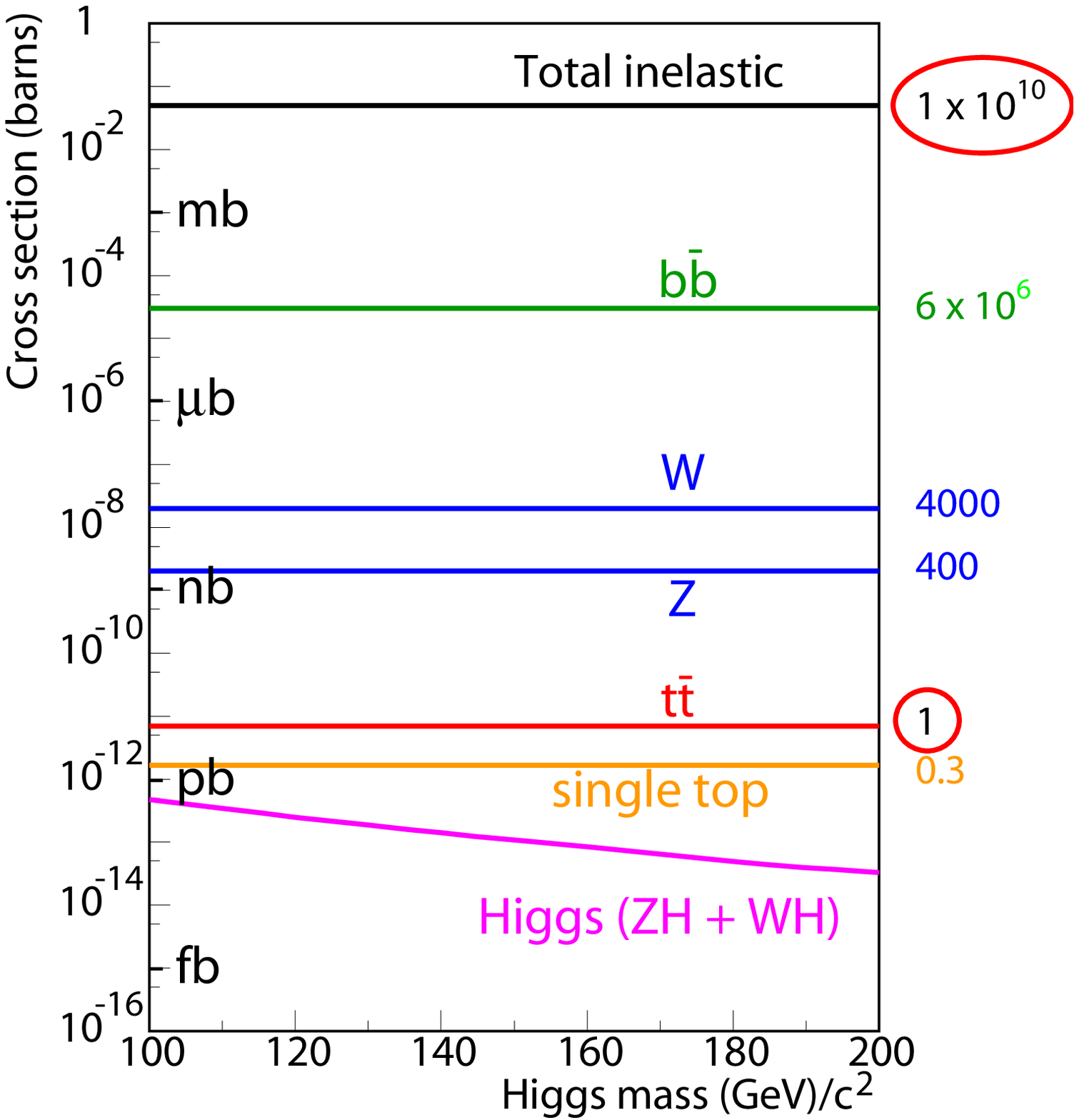} 
 \caption{Cross-sections for  processes that contribute to the background in
  $t\bar t$ cross-section measurements.\label{fig:xsec_range}
}
\end{minipage}
\end{figure}

At the Tevatron top quarks are most likely produced in pairs.
The dominant process is  quark anti-quark annihilation to a gluon
that then splits into $t\bar t$ (\fig{fig:ttbar_feynman}). 
A pair production cross-section is thus probing our understanding of the
strong force and its couplings to the top quark.

The top-quarks produced subsequently decay to a $b$-quarks and a $W$-boson to nearly 100\%.
Decay modes  of $t\bar t$ events are thus determined by the decay modes of the
$W$s. Dilepton events feature two jets from $b$-quarks, two leptons and a
transverse momentum imbalance (missing transverse energies, $\met$) stemming
from the 2 neutrinos that escape the detector.
Lepton plus jets events consists of two jets from $b$-quarks, two additional
jets from light quarks, a single lepton and missing transverse energy from the
neutrino.
The alljets events will have at least 6 jets, 2 of which stem from $b$-quarks.
As taus in the final state are difficult to identify usually only electron
and muons are used in the lepton channels.
The dilepton, lepton plus jets and alljets then contribute with
$5\%$, $30\%$ and $46\%$, respectively.

In order to measure a cross-section it is necessary to count the signal
events in the data. In addition it is important to understand the amount of
background events which remain after selection.
In $t\bar t$ events background arises from multijet events, $W$- and $Z$-production.
In multijet events instead of a pair of top quarks a pair of light (or $b$-)quarks is
produced. Additional jets arise from initial and final state gluon radiation.
$W$- and $Z$-bosons are also produced by $q\bar q$ annihilation. Their
leptonic decay modes can look like signal when additional jet arises from gluon radiation.

Besides these physics backgrounds instrumental backgrounds are important.
Misidentification of physics objects as being leptons and momentum
mismeasurement leading to overestimated $\met$ are the most important.
These misidentification and mismeasurements are in themselves rare, however,
the cross-sections for the backgrounds are orders of magnitude higher than
that for $t\bar t$ production (see \fig{fig:xsec_range}).

\begin{figure}[tbp]
  \centering
  \includegraphics[width=0.48\textwidth]{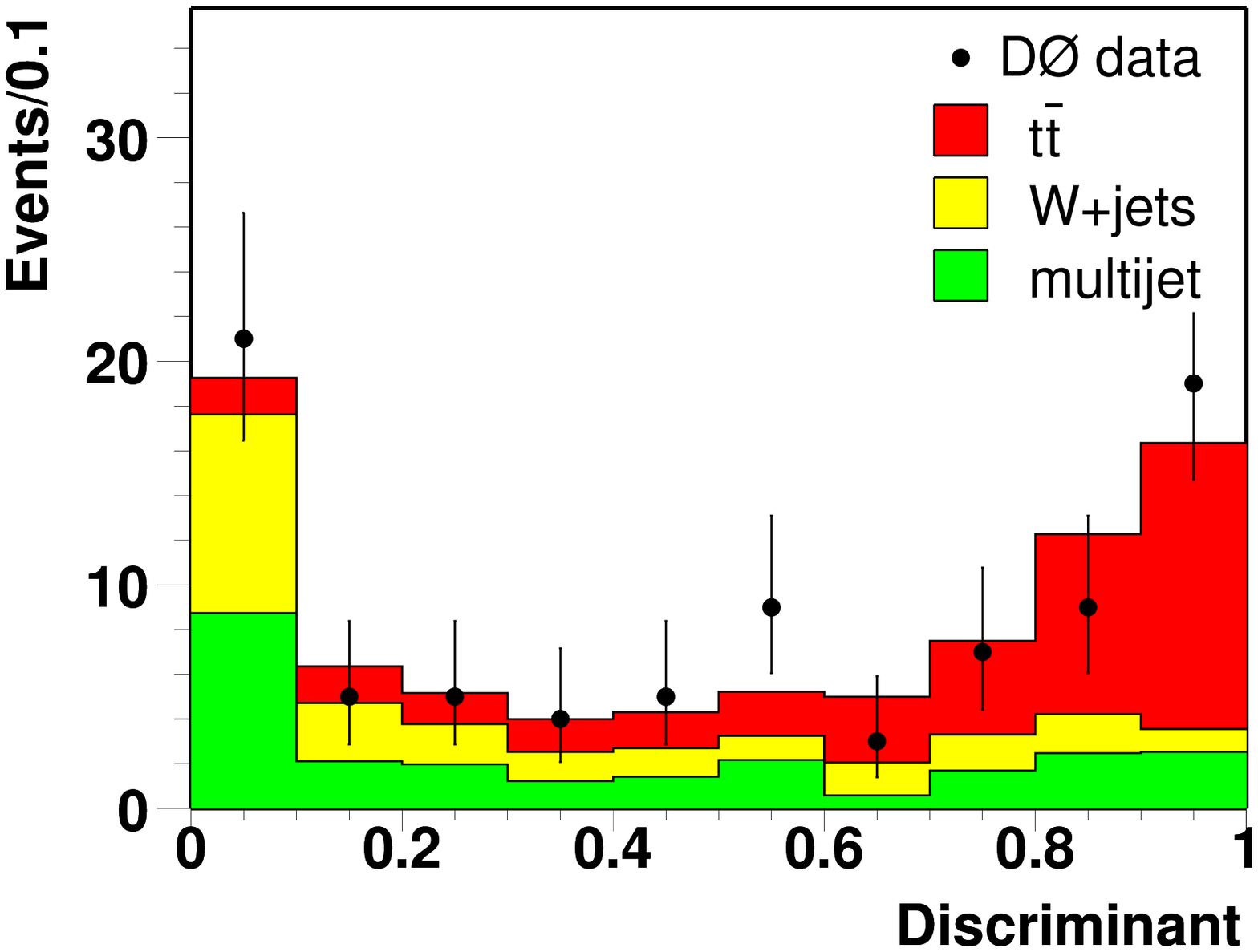}
  \includegraphics[width=0.48\textwidth]{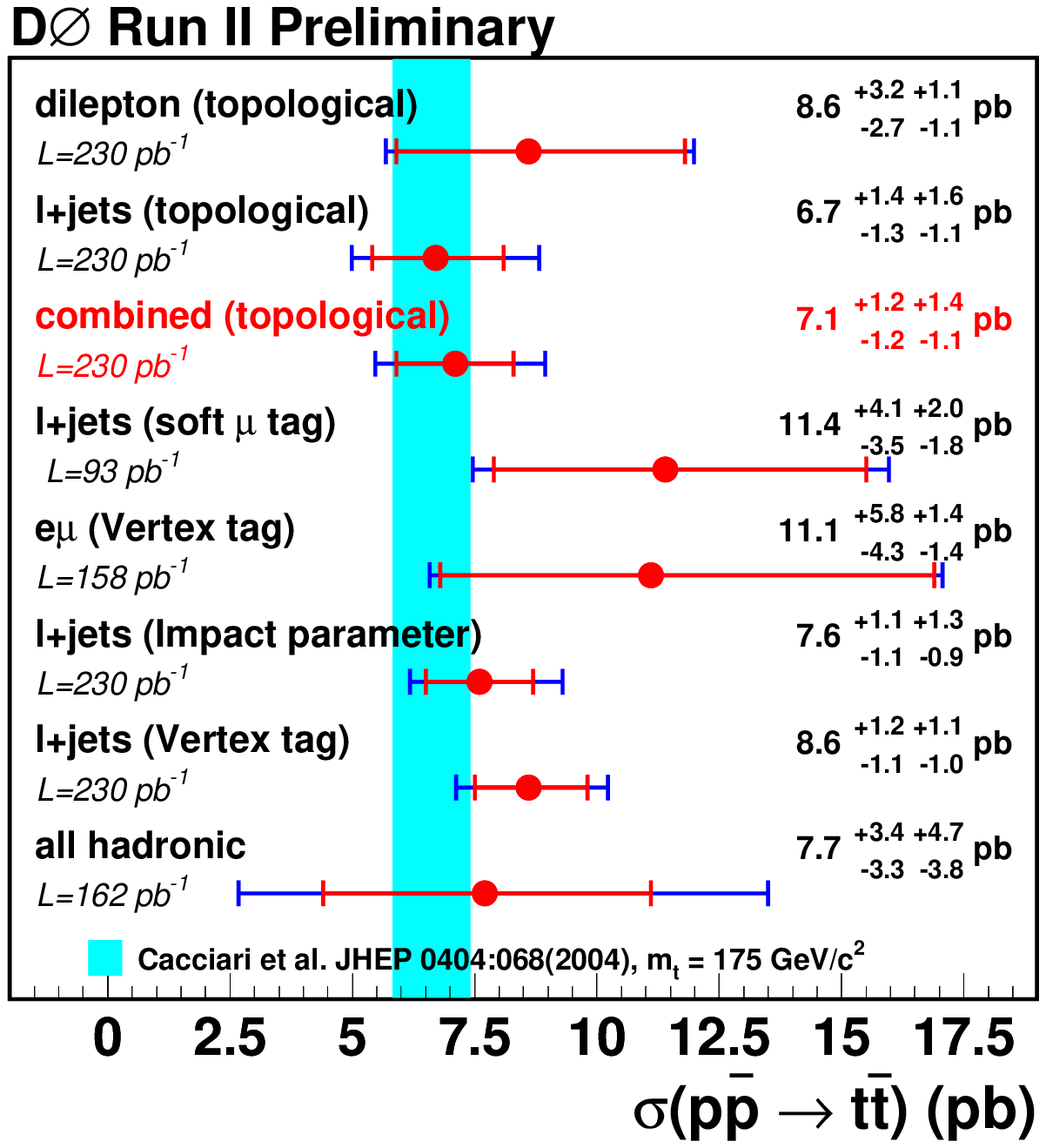}
 \caption{Left: Topological discriminant used for statistical separation of 
  $t\bar t$ signal from background with the expected background and signal
  contributions. Right: Comparison of D\O\ Run II results on the top pair
  production cross section for various methods and 
  channels~\cite{Abazov:2005yt,Abazov:2005ex,d0note4528,Abazov:2005ey,d0note4428}
  compared to the Standard Model expectation~\cite{Berger:1997gz,Bonciani:1998vc,Kidonakis:2003qe,Cacciari:2003fi}.}
\label{fig:ttbar_xsec}
\label{fig:ttbar_discriminant}
\end{figure}

All analyses start by selecting event by event the signatures expected from
the relevant final state. In the lepton plus jets $t\bar t$ analysis, which shall serve
as an example here, this means
requiring at least $4$ jets, an isolated (non-collinear) lepton and missing
transverse energy.  All objects are required to have a transverse momentum
larger than $20\GeV$.
This yields 87 $e+$jets and 80 $\mu+$jets events in $230\ipb$.

The efficiency of the selection is determined from applying this selection to 
simulated signal events. The agreement between simulation and data was checked
in various distributions at preselection level. Additional smearing was
applied where necessary. The efficiencies obtained are $(11.6\pm1.7)\%$ and
$(11.7\pm1.9)\%$ for the $e+$jets and $\mu+$jets channel, respectively.

The background within the selected samples is dominated by $W+$jets events,
which have the same signature as $t\bar t$ events. The samples also include
contribution from multijet events from instrumental background.
In order to determine the amount of background two method are applied.
The instrumental background is taken from data following the ``matrix''
method described in~\cite{Abbott:1999tt}.

To estimate the physical $W+$jets background a discriminant is built that allows 
separating the signal from background on a statistical basis.
The optimal discriminant was found to be built from six observables:
{\it i)} $H_T$, the scalar sum of the $p_T$ of the four leading jets;
{\it ii)} $\Delta\phi(l,\met)$, the azimuthal opening angle between the
lepton and the missing transverse energy;
{\it iii)} $K_{T\mathrm{min}}$,  the minimum of an energy normalised distance
between pairs of jets;
{\it iv)} ${\cal C}$, the centrality;
{\it v)} ${\cal A}$, the event aplanarity;
{\it vi)} ${\cal S}$, the event sphericity.

The amount of signal and $W+$jets background is then fitted to the observed
distribution using the shapes expected from simulation. 
The instrumental
background is kept fixed to the amount obtained from the matrix method. 
The resulting composition of signal and background is visualised in
\fig{fig:ttbar_discriminant} as function of the events discriminant value.
The final result from applying this method to $230\ipb$ of data~\cite{Abazov:2005ex} is
\begin{equation}
\sigma_{t\bar t}=6.7^{+1.4}_{-1.3}\null_\mathrm{stat}\null^{+1.6}_{-1.1}\null_\mathrm{syst}\pm0.4_\mathrm{lumi}\,\pb
\end{equation}
The result agree with the Standard Model prediction of 
$\sigma_{t\bar t}=6.77\pm0.42\pb$~\cite{Berger:1997gz,Bonciani:1998vc,Kidonakis:2003qe,Cacciari:2003fi}. 
It is also consistent with results obtained with other methods and from other
channels as presented in \fig{fig:ttbar_xsec}.
Thus at the current level of precision there is no hint of deviations from the
Standard Model, neither regarding the amount of production nor regarding the
composition of decay channels.

\section{Single top production}

Beside being produced via strong interaction the top can also be produced via
the weak interaction. When the intermediate boson is a $W$ it is produced singly.
Measuring the single top production cross-section thus tests the strength of
the coupling at the $Wtb$-vertex. 

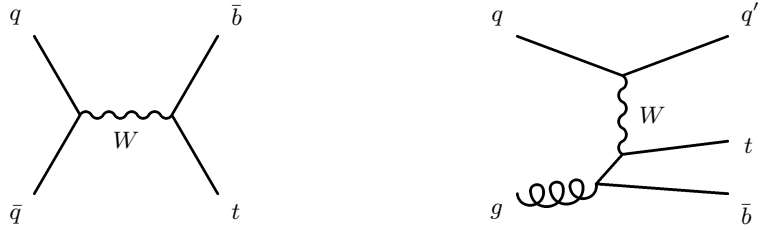
\begin{figure}[t]
\vspace*{1em}
  \begin{fmffile}{mf_single_top}
\null\hfill
    \begin{minipage}[c]{35mm} \footnotesize
    \setlength{\unitlength}{0.7mm}
    \begin{fmfgraph*}(35,30)
      \fmfstraight
      \fmfleft{em,ep}
      \fmfright{q,qb}
      \fmf{plain}{em,v,ep}
      \fmf{photon,lab=\raisebox{-2mm}{$W$}}{v,vv}
      \fmf{plain}{q,vv,qb}
      \fmflabel{$q$}{ep}
      \fmflabel{$\bar{q}$}{em}
      \fmflabel{$t$}{q}
      \fmflabel{$\bar{b}$}{qb}
    \end{fmfgraph*}
   \end{minipage}\hfill
    \begin{minipage}[c]{35mm} \footnotesize
    \setlength{\unitlength}{0.7mm}
    \begin{fmfgraph*}(40,30)
      \fmfstraight
      \fmfleft{g1,q1}
      \fmfright{b,t,dummy,q2}
      \fmf{plain}{q1,vv,q2}
      \fmf{phantom}{g1,dv,b}
      \fmf{phantom}{vv,dv}
      \fmffreeze
      \fmf{gluon,tension=2}{g1,g2}
      \fmf{plain}{t,dv,g2,b}
      \fmffreeze
      \fmf{photon,lab={$W$}}{dv,vv}
      \fmflabel{$g$}{g1}
      \fmflabel{$q$}{q1}
      \fmflabel{$q'$}{q2}
      \fmflabel{$t$}{t}
      \fmflabel{$\bar{b}$}{b}
    \end{fmfgraph*}
   \end{minipage}
\hfill\null
\vspace*{1em}
\end{fmffile}

 \caption{Feynman diagrams for single top production. Left:
   $s$-channel. Right: $t$-channel.}
\label{fig:singletop_feynman}
\end{figure}

Two different Feynman diagrams with different final states 
contribute to the single top production, an $s$-channel diagram in which the top is
accompanied by a $b$-quark and a $t$-channel diagram in which it is accompanied by a
light quark and a $b$-quark
diagram, see \fig{fig:singletop_feynman}.

\begin{figure}[tbp]
  \centering
\begin{minipage}[b]{0.48\textwidth}
  \includegraphics[width=\textwidth]{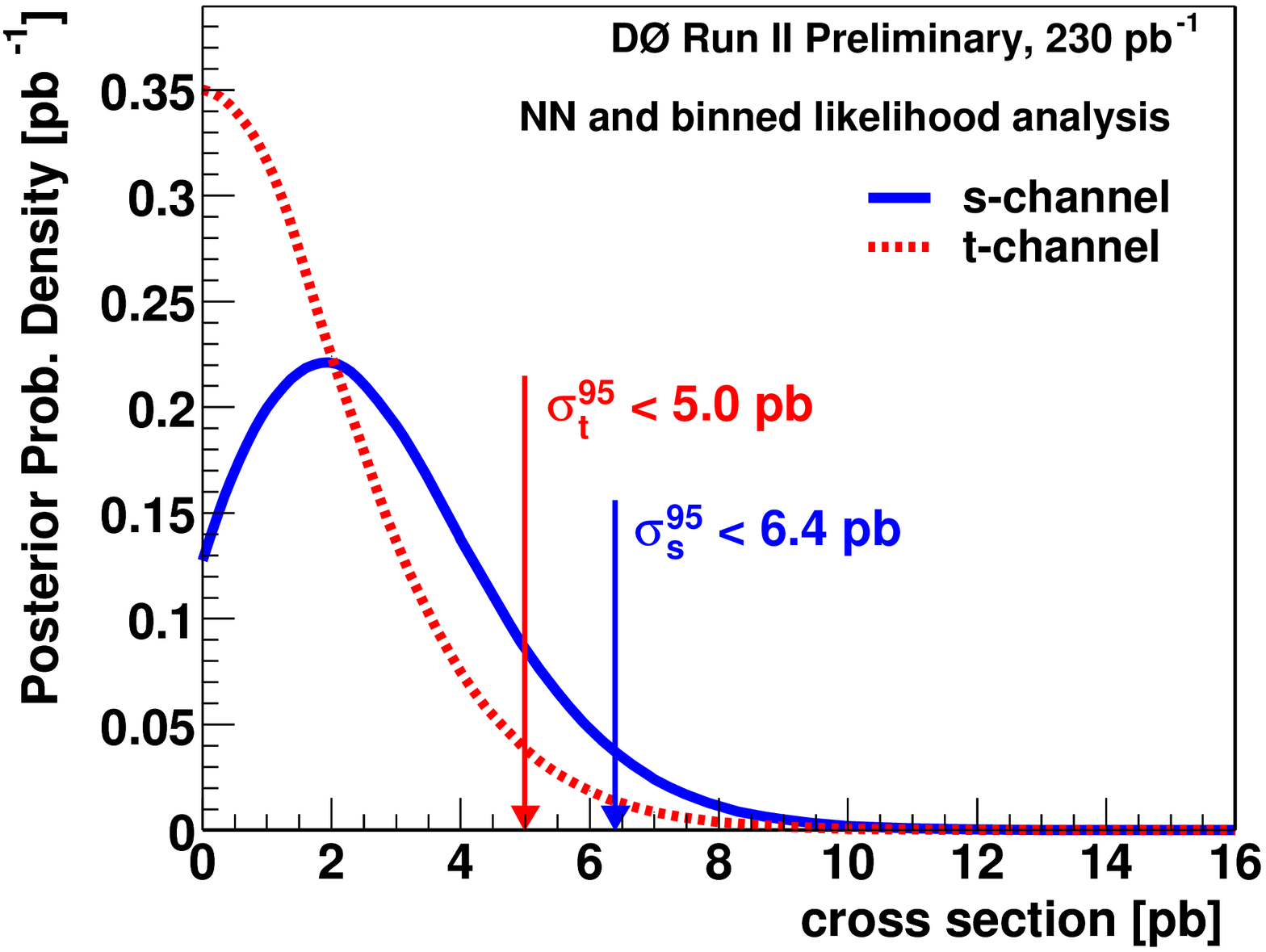}
 \caption{Posterior probability density for $s$- and $t$-channel.}
\label{fig:singletop_probdens}
\end{minipage}\hfill
\begin{minipage}[b]{0.48\textwidth}
  \includegraphics[width=\textwidth]{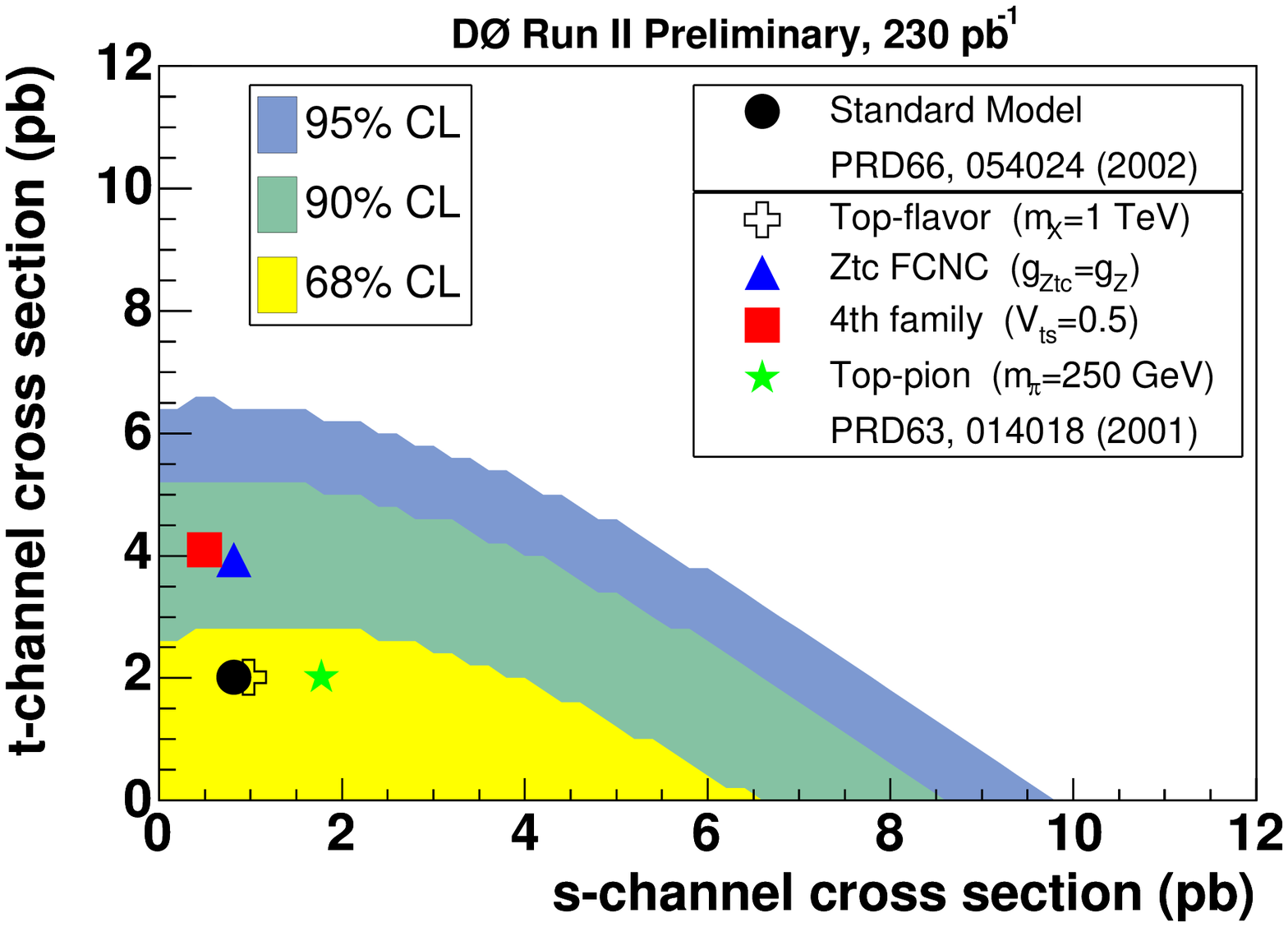}
  \caption{2-dimensional exclusion limits of single top production obtained from a split dataset compared to
  various extensions of the Standard Model.}
\label{fig:singletop_2dlimit}
\end{minipage}
\end{figure}

For the current analyses only the leptonic decay mode of the $W$ stemming from
the top-decay is considered. 
While for the $s$-channel the $b$-quark produced together with the top quark is expected to be
visible in the detector, for the $t$-channel the associated $b$-quark is
likely to disappear along the beam pipe. So only a light quark with large
transverse momentum is expected in addition to the top decay products.

In order to perform this analysis efficiently it is important to distinguish
between light quark jets  and  $b$-quark jets. To tag jets that stem from a
$b$-quark the long $b$-lifetime is exploited. Hadrons formed from $b$-quarks
can travel several millimeters before they decay.

One method of $b$-tagging is through reconstruction of secondary vertices.
Tracks associated with a $B$-hadron decay will form a vertex which due to the long
lifetime is separated from the primary interaction vertex.
The  displacement of the secondary vertex is used to
identify $B$-hadrons. The tagging efficiency and purities directly enter into
the signal efficiencies and purities for the single top analysis.
Thus this analysis is specifically profiting from the improved tracking that
is needed for a precise vertex reconstruction and has been provided by the 2003 data reprocessing.

The actual single top analysis is performed separately for the two channels,
treating muons and electrons and further single and double tagged events
separately. For each of these 8 analysis chains a neural network is trained to
distinguish signal from background.

The final results are obtained from a binned log-likelihood. In $230\ipb$ of
data no excess over the expected background is observed and upper limits on the
cross-sections are set~\cite{Abazov:2005zz}:
\begin{equation}
\mbox{s-channel:}\quad \sigma < 6.4 \pb \quad 95\%\mathrm{CL}\qquad\qquad
\mbox{t-channel:}\quad \sigma <5.0 \pb \quad 95\%\mathrm{CL}
\end{equation}
\fig{fig:singletop_probdens} shows the posterior probability densities.
These are the current worlds best limits. In a two dimensional presentation 
of these cross-section limits in \fig{fig:singletop_2dlimit} one can
see that these limit  are starting to reach the interesting
region. Several extensions of the Standard Model can be checked before the
sensitivity reaches the level of the Standard Model expectation.

\section{Top quark mass measurement}
\label{sec:topmass} 
The mass of the top quark as with all other fermion masses, isn't predicted by the Standard Model. 
However, they enter virtual corrections to various processes. A precise
knowledge of the top mass can thus still be used for checking the consistency
of the Standard Model. It is also used to restrict the masses allowed
for the Higgs boson within the Standard Model.

Here the most recent result on the top quark mass obtained be D\O\ is
reported. It is based on the semileptonic decay channel with a selection
identical to the one described for the cross-section measurement above.
After the initial selection the sample is purified by cutting on a
discriminant similar to the one used in that measurement.

\begin{figure}[b]
  \centering
\begin{minipage}[t]{0.48\textwidth}
  \includegraphics[width=\textwidth]{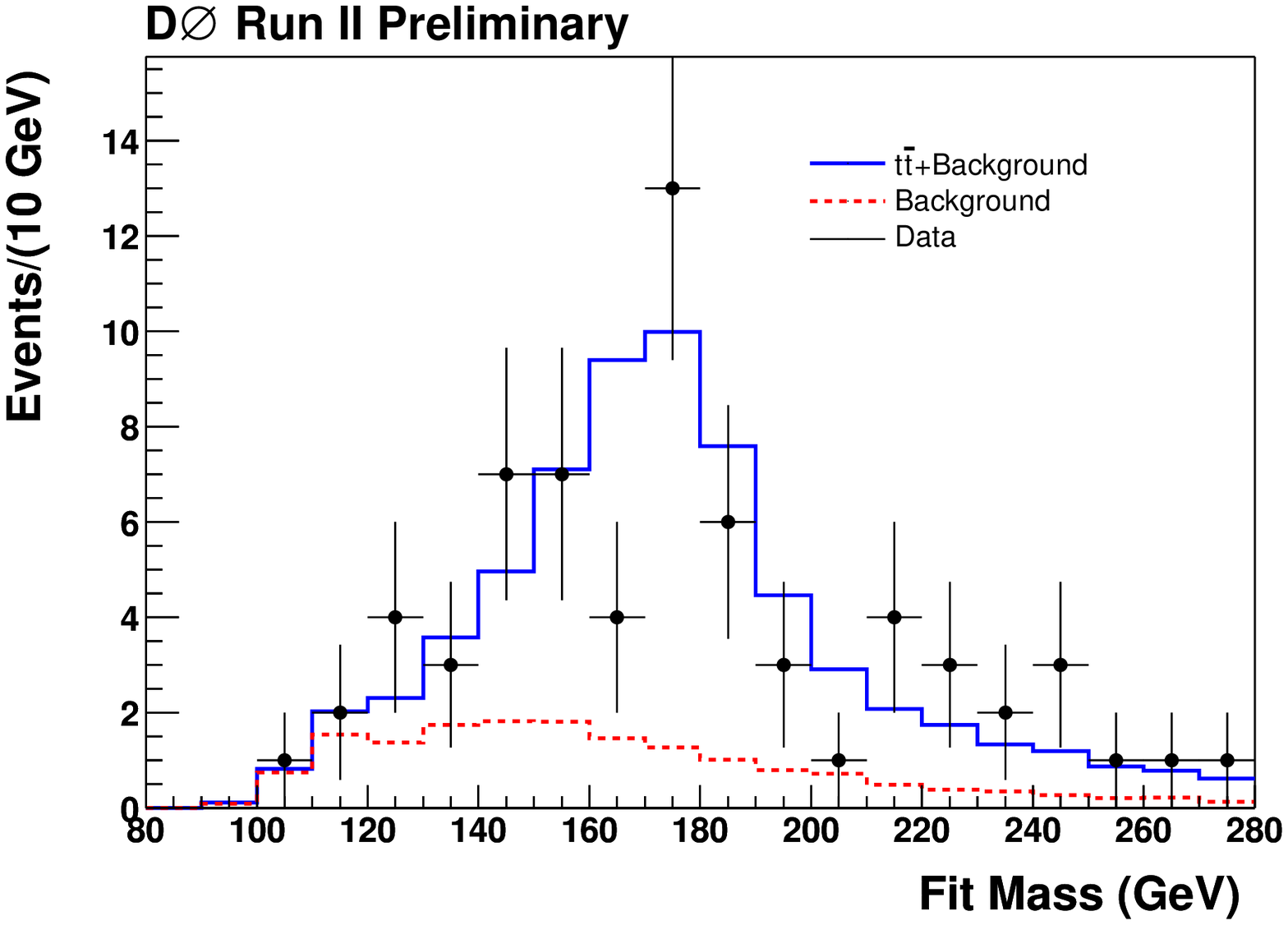} 
 \caption{Reconstructed top masses for $b$-tagged events compared to the
  simulation of signal and background. The signal simulation with $m_t$
  closest to the final result is shown.}
\label{fig:mtop_template}
\end{minipage}
\hfill
\begin{minipage}[t]{0.48\textwidth}
  \includegraphics[width=\textwidth]{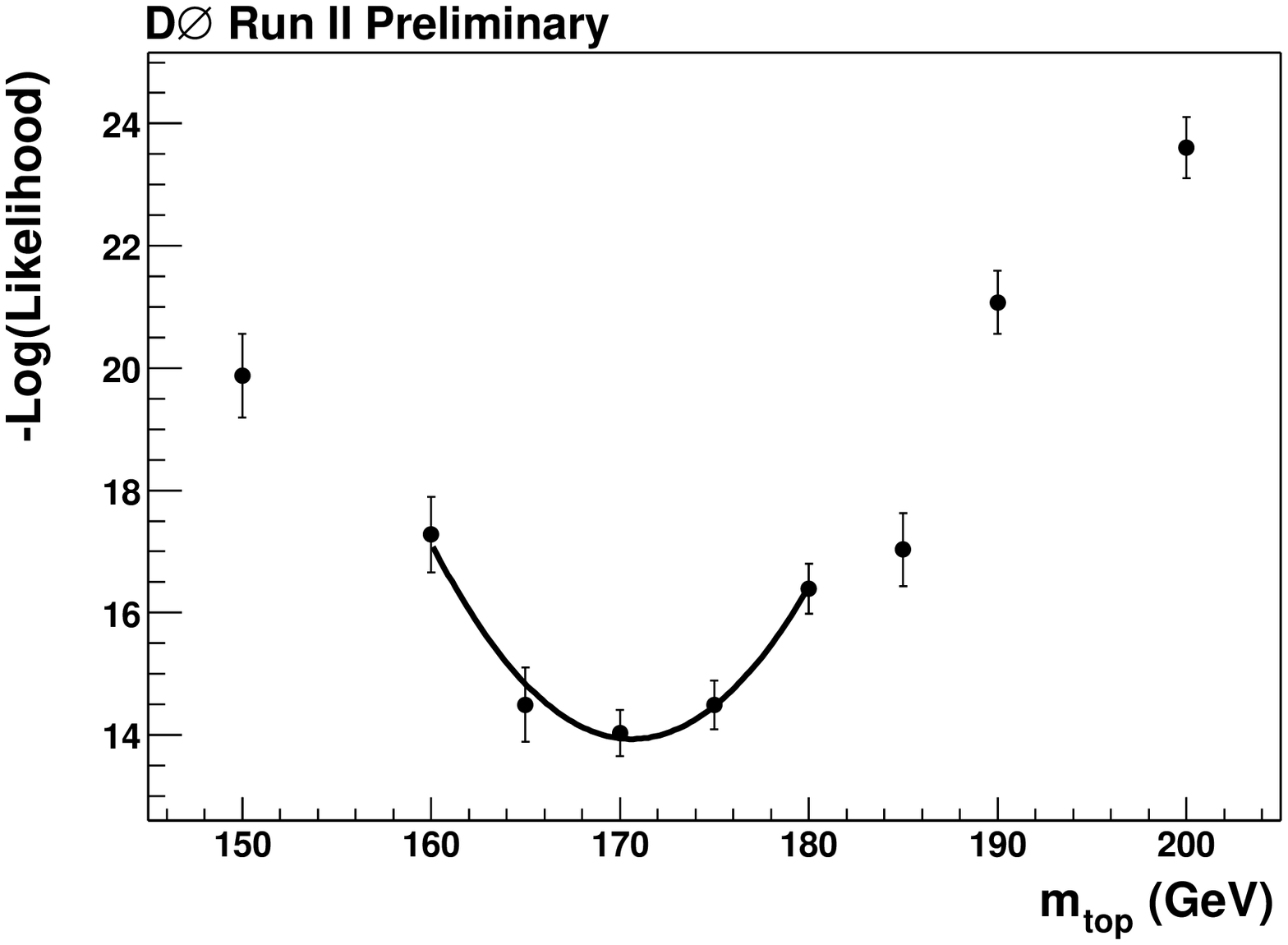}
 \caption{The likelihood curve from the fit of $b$ tagged events to 
templates with varying top mass in the signal simulation.}
\label{fig:mtop_likelihood}
\end{minipage}
\end{figure}
\begin{figure}
\begin{minipage}[b]{\textwidth}
\centering
  \includegraphics[width=0.48\textwidth]{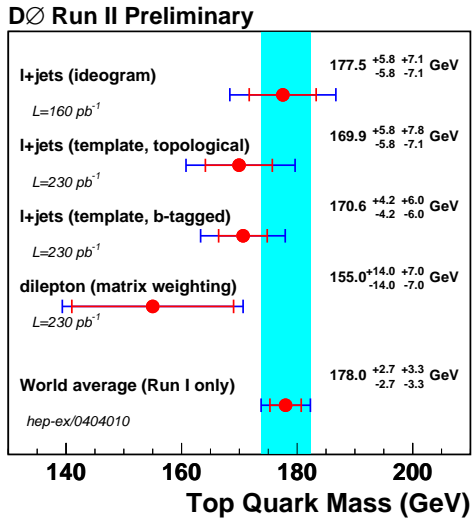}
 \caption{D\O\ Run~II results on the top quark mass obtained from various
  methods and channels~\cite{d0note4574,TopMassTemplate2005,d0note4725} 
 compared the Run~I world average~\cite{Azzi:2004rc}.}
\label{fig:mtop_summary}
\end{minipage}
\end{figure}

Then the mass of the top quark is reconstructed event by event.
First the full neutrino momentum needs to be recovered.
The missing $z$-coordinate is recovered by requiring that 
the combined invariant mass of particles assumed to be from the $W$ boson 
is consistent with the $W$-mass.
Then the lepton-neutrino pair and the four jets need to be assigned
to the two top quarks.
The invariant mass of these two triples is the reconstructed top quark mass.
The correct assignment of particles to their parents 
as well as precisely measured momenta are crucial for
this measurement.

To determine the overall result the top masses reconstructed event-by-event
are filled into a histogram and compared to simulations that were done for
various hypothetical values of $m_t$ and that include the expected amount of
background (\fig{fig:mtop_template}).
Finally the best value for $m_t$ is obtained with a maximum likelihood method
 (\fig{fig:mtop_likelihood}).

Results are presented for two options. A purely topological selection
that uses all the described data and a selection which requires at least one
$b$-tag thereby reducing the number of possible assignments of jets to the
top-quarks and also reducing the background~\cite{TopMassTemplate2005}: 
\begin{eqnarray}
m_t&=&169.0\pm 5.8_\mathrm{stat}\null^{+7.8}_{-7.1}\null_\mathrm{syst}\GeV\,\,\,\,\,\qquad\mbox{(topological)}\nonumber\\
m_t&=&170.6\pm 4.2_\mathrm{stat}\pm 6.0\null_\mathrm{syst}\GeV\qquad\mbox{($b$-tagged)}
\end{eqnarray}

In \fig{fig:mtop_summary} this result is compared to other results of $m_t$ in
D\O. All method are consistent with each other and also with the result
obtained in Run~I.

While the measurements of the top quark mass rely on simulation and a good
tracking for an efficient $b$-tagging, they also depend on the overall
calibration of the calorimeters to determine the measured jet energies.

This last issue has been tackled by improving the calibration of the
calorimeter. This new calibration is applied during the currently ongoing
data-reprocessing (see Section \ref{sect:d0repro}). 
Added to the amount of new data expected from the Tevatron during this year
the reprocessing effort will again be responsible for doubling the
dataset available for coherent physics analysis in 2006.

\section{Summary}

With the example of three top analyses it has been shown how  physics 
results rely on D\O's ability of successful Grid computing.

Distributed computing has been used for production of simulated events since
the beginning of Tevatron Run II. In addition re-reconstruction of data in
order to apply improved reconstruction algorithms or calibration constants 
is currently being performed in a distributed grid like manner for the second time.
D\O\ is  relying on their SamGrid project to perform these tasks.

Physics analyses rely on simulation 
and profit from improved algorithms implemented with improved detector
understanding:
the described top pair production cross-section measurement relies on the 
ability to extract the shape of a discriminant in signal and background
simulation  in order to determine the signal to background ratio in the
selected events. $b$ tagging, which is a very powerful technique for enhancing the
signal to background ration in top events, profits from the tracking
improvements that were made available through the first reprocessing.
D\O's single top cross-section limits which are currently the worlds best limits
are directly dependent on the $b$-tagging efficiency.
The top mass measurement in addition requires an excellent understanding of
the jet energy scale. The current data re-reconstruction effort is applying
improved calorimeter calibration constants which will reduce the systematic
uncertainty due to the jet energy scale, which currently dominates the uncertainty.

These top quark measurements are just examples that illustrate how the production of
simulated events impacts the physics results and
where it is important to increase the dataset reconstructed with the
latest algorithms by re-reconstructing older data.
Other analyses in D\O's wide physics programme profit from these efforts in a similar manner. Besides top
quark measurements this include  all aspects of the
Standard Model as well as direct searches for new physics.
Grid computing enables D\O\ to get the most of these data.


\section*{References}
\bibliographystyle{utphys}
\begin{flushleft}
\bibliography{Grid,Dzero}
\end{flushleft}
\end{document}